\documentclass[journal=ancac3 manuscript=article]{achemso}
\setkeys{acs}{etalmode=truncate}
\setkeys{acs}{maxauthors=10}

\title{Influence of Chemisorbed Oxygen on the Growth of Graphene on Cu(100) by Chemical Vapor Deposition}

\author{Zachary R. Robinson}
\affiliation{College of Nanoscale Science and Engineering, University at Albany - SUNY}
\alsoaffiliation{U.S. Naval Research Laboratory, Washington, DC}
\author{Eng Wen Ong}
\author{Tyler R. Mowll}
\author{Parul Tyagi}
\affiliation{College of Nanoscale Science and Engineering, University at Albany - SUNY}
\author{D. Kurt Gaskill}
\affiliation{U.S. Naval Research Laboratory, Washington, DC}
\author{Heike Geisler}
\affiliation{Department of Chemistry and Biochemistry, SUNY College at Oneonta}
\author{Carl A. Ventrice, Jr.}
\email{CVentrice@albany.edu}
\phone{518 437 - 8693}
\fax{518 437 - 8687}
\affiliation{College of Nanoscale Science and Engineering, University at Albany - SUNY}

\begin{document}

\begin{abstract}
Understanding the influence that copper substrate surface symmetry and oxygen impurities have on the growth of graphene by chemical vapor deposition is important for developing techniques for producing high quality graphene.  Therefore, we have studied the growth of graphene by catalytic decomposition of ethylene in an ultra-high vacuum chamber on both a clean Cu(100) surface and a Cu(100) surface pre-dosed with a layer of chemisorbed oxygen. The crystal structure of the graphene films was characterized with \emph{in-situ} low energy electron diffraction. By heating the clean Cu(100) substrate from room temperature to the growth temperature in ethylene, epitaxial graphene films were formed.  The crystal quality was found to depend strongly on the growth temperature.  At 900~$^\circ$C, well-ordered two-domain graphene films were formed. Pre-dosing the Cu(100) surface with a chemisorbed layer of oxygen before graphene growth was found to adversely affect the crystal quality of the graphene overlayer by inducing a much higher degree of rotational disorder of the graphene grains with respect to the Cu(100) substrate.  The growth morphology of the graphene islands during the initial stages of nucleation was monitored with \emph{ex-situ} scanning electron microscopy.  The nucleation rate of the graphene islands was observed to drop by an order of magnitude by pre-dosing the Cu(100) surface with a chemisorbed oxygen layer before growth.  This reduction in nucleation rate results in the formation of much larger graphene islands.  Therefore, the presence of oxygen during graphene growth affects both the relative orientation and average size of grains within the films grown on Cu(100) substrates.


\end{abstract}

\section{Introduction}
Graphene, which is a single layer of sp$^2$ bonded carbon atoms, shows promise for use in a variety of technological applications. For instance, it has an optical transmittance of over 97\% throughout the visible range and is a semi-metal, which makes it an ideal material for use as a transparent conducting medium for optical displays or solar cells \cite{wang2008transparent, 5347313820100801}. Its compatibility with conventional Si-based processing techniques and its intrinsic mobility of over 200,000 cm$^2$/V$\cdot$s make graphene a potential candidate for use in high speed electronic devices \cite{Bolotin2008351, Lin10062011}. In addition, the sheet thermal conductivity of graphene has been measured to be 5 $\times$ 10$^3$ W/m$\cdot$K, which is considerably higher than the thermal conductivity of bulk Cu (400~W/m$\cdot$K) \cite{doi:10.1021/nl0731872}. This ability of graphene to efficiently dissipate heat is especially important if it is to be implemented in high-density integrated circuits.

Although graphene shows great promise for use in several technological applications, it has not seen widespread adoption within the semiconductor industry.  The primary reason for this is that a low-cost technique for producing large areas of graphene with a defect density that is low enough for most applications has not yet been developed.  Although techniques for producing high quality graphene on SiC substrates have been developed, the cost of the SiC substrates is prohibitively high. Chemical vapor deposition (CVD) of graphene on copper foil substrates is one of the most promising techniques for producing large area graphene films since the substrate cost is relatively low and the graphene thickness is self-limited to a single layer of graphene when low pressure conditions are used \cite{Li05062009}. However, controlling the defect density of graphene grown by CVD has been a challenge.  Because of this, a wide range of carrier mobilities have been reported for graphene films grown by CVD on Cu foil substrates \cite{C0JM02126A, guermoune2011chemical}.  Imperfections in the graphene films can result from several different sources.  For instance, graphene must be transferred from the Cu substrate to an insulating or semiconducting substrate for characterization of its mobility, which can result in structural damage during the transfer process.  Chemical residues from the etching of the Cu substrate will also adversely affect the transport properties.  However, most of the variability of the transport properties of graphene grown by CVD most likely results from the wide range of vacuum conditions, temperature profiles, and gas purities used during the growth process \cite{C0JM02126A}. For growth conditions that produce large graphene grains with a low number of structural defects and impurities, the transport properties have been shown to approach those of graphene mechanically exfoliated from graphite \cite{doi:10.1021/nl101629g}.  On the other hand, graphene films composed of small grains that are rotationally misaligned with respect to each other are expected to have a reduced mobility and thermal conductivity owing to the scattering of carriers and phonons at grain boundaries.  In addition, structural imperfections at grain boundaries and within the grain will have a much higher reactivity towards atmospheric contaminants.  This will result in a time dependent degradation of the transport properties.

There are several factors that will affect the graphene grain size and defect density during the growth by CVD.  For instance, the size and orientation of the Cu substrate grains are expected to influence the rotational orientation and size of the graphene grains.  Since the interaction of graphene and Cu is weak, the rotational orientation of the graphene grains during the initial nucleation is expected to be preserved as the graphene grains grow laterally.  Step edges and point defects at the surface of the Cu substrate typically have a much higher catalytic activity towards the decomposition of the hydrocarbon precursor and a higher interaction strength with carbon atoms.  Therefore, these sites are expected to be the primary nucleation sites for graphene at lower temperatures.  At higher temperatures, the interaction of the carbon atoms with the ordered sites on the terraces between the Cu substrate step edges should be comparable to the interaction at the step edges.  Once this happens, nucleation on the terraces is expected to dominate.  Since the terrace sites have a well-defined symmetry, the graphene grains will typically have a few preferred orientations with respect to the substrate.

The two lowest energy surface terminations of Cu, which is a face-centered cubic (FCC) crystal, are the (111) and (100) orientations.  For cold-rolled foils of Cu, the grains within the Cu are known to predominantly reorient with a (100) surface termination at the elevated temperatures used during CVD growth of graphene \cite{doi:10.1021/ja109793s, robinson:011401}.  However, thin films of Cu deposited on substrates such as sapphire or SiO$_2$ often have a (111) texture \cite{citeulike:9349235}. Because both graphene and the Cu(111) surface have hexagonal symmetry and the lattice mismatch between them is only -3.5\%, it should be possible to grow overlayers of graphene on this surface with a single rotational orientation if nucleation occurs on the substrate terraces.  In fact, previous studies of graphene growth in ultra-high vacuum (UHV) on Cu(111) single crystals have shown that single-domain epitaxial films of graphene can be grown \cite{PhysRevB.84.155425, robinson2012argon}. Because the (100) surface has square symmetry, whereas the graphene lattice is hexagonal, graphene grown on this surface will have a minimum of two preferred orientations.  Somewhat conflicting results have been reported for the rotational alignment of graphene films on Cu(100) single crystals.  Rasool \emph{et al}. used scanning tunneling microscopy (STM) to measure the atomic structure of graphene films grown in a tube furnace on Cu(100) single crystals and found that the graphene grains had nucleated in a wide range of orientations on this surface \cite{doi:10.1021/ja200245p}.  In contrast to those results, Zhao \emph{et al}. used \emph{in-situ} STM and low energy electon diffraction (LEED) to analyze the crystal struture of graphene films grown on a Cu(100) single-crystal in their UHV system and reported that a two-domain overlayer had formed \cite{Zhao2011509}.  However, the presence of broad arcs instead of spots in their LEED patterns indicates that the graphene grains within their films also had considerable rotational disorder.  Woford \emph{et al}. used \emph{in-situ} low energy electron microscopy (LEEM) to analyze the crystal structure of graphene grains grown by carbon deposition in UHV on Cu foil substrates that had recrystallized with a (100) texture and found a similar result: a two-domain overlayer with a mosaic spread of $\pm 7.5^\circ$ \cite{doi:10.1021/nl102788f}.  Therefore, it is still an open question as to whether it is possible to grow a graphene overlayer where each domain is rotationally aligned within a few degrees of one of the two high symmetry directions of the substrate.

As mentioned above, one of the most popular substrates for graphene growth are cold-rolled Cu foils.  For growth of graphene by CVD on foils, an important factor that can influence the growth rate and the defect density of graphene is the presence of copper oxide within the foil and oxygen containing molecules such as O$_2$ and H$_2$O in the chamber and/or gas stream during growth.  The copper oxide can potentially affect both the growth rate of the grains within the Cu foil and the catalytic activity of the surface.  Oxygen in the gas stream will etch the graphene at elevated temperatures by forming CO$_2$ and CO.  During conventional CVD growth of graphene, the Cu foil substrates are annealed in flowing hydrogen before graphene growth to reduce the copper oxide on the surface and within the bulk of the foil.  The annealing process also results in an increase in the size of the grains within the Cu substrate.  To initiate graphene growth, a hydrocarbon precursor gas is introduced into the growth chamber, usually mixed with hydrogen to prevent copper oxide formation during the graphene growth process.  After the growth process is complete, the sample is cooled to room temperature in a flow of the hydrocarbon precursor mixed with hydrogen.  The most common chambers used for graphene growth are hot walled reactors (\emph{i.e.}, tube furnaces).  For atmospheric pressure CVD reactors, a flow of an inert gas such as argon is often used to maintain low levels of impurity gases in the reactor and to dilute the hydrocarbon precursor.  Low pressure CVD reactors are generally cleaner than atmospheric pressure reactors.  However, the base pressure of most of these systems is in the mTorr region, so these systems are still far from the purity that can be achieved with UHV-based growth chambers.

One of the first studies of the influence of oxygen on the rate at which a hydrocarbon precursor decomposes on a Cu single crystal substrate was published by Alstrup \emph{et al}. in 1992 \cite{Alstrup199295}. In this study, the decomposition of methane (CH$_4$) on both the clean and oxygen pre-dosed Cu(100) surface was performed in a UHV chamber.  The decomposition rate  of the methane was monitored with \emph{in-situ} X-ray photoelectron spectroscopy (XPS).  Because of the exceedingly low dissociative chemisorption probability, extreme care was taken to reduce impurities in the methane gas by using a bakeable gas inlet system that incorporated a cooled molecular sieve and a nickel catalyst.  The decomposition studies were performed with a methane pressure of 10 Torr over the temperature range of 800-1000 K for the clean surface and 700-800 K for the oxygen pre-dosed surface.  The oxygen pre-dosed surface was prepared by exposing the surface to 1000 L of molecular oxygen at 500 K, which resulted in a saturation coverage of 0.5 monolayers (ML) of chemisorbed oxygen on the surface.  It was determined that the activation energy for dissociative adsorption was 201 $\pm 4$ kJ/mol and 123 $\pm 27$ kJ/mol on the clean surface and the oxygen pre-dosed surface, respectively.  Interestingly, they also determined that the saturation coverage of carbon on the Cu(100) surface was $\sim$2.4~ML, which corresponds to a single atomic layer of graphene ($\sigma _{Cu(100)} = 1.53 \times 10^{15}$ cm$^{-2}$ and $\sigma _{graphene} = 3.82 \times 10^{15}$ cm$^{-2}$).  Therefore, they were apparently the first group to determine that graphene growth self-terminates at a single atomic layer on a Cu substrate when using a methane precursor.

In order to study the effect that oxygen has on the growth of graphene on Cu substrates, we have grown graphene films in an UHV chamber by catalytic decomposition of ethylene (C$_2$H$_4$) on a Cu(100) single crystal substrate with and without a chemisorbed oxygen layer on the surface and monitored the crystal structure of the graphene with \emph{in-situ} LEED and the growth morphology with \emph{ex-situ} scanning electron microscopy (SEM).  The Cu(100) surface was chosen because the grains within Cu foils used for graphene growth typically recrystallize with a (100) surface termination.  By performing this study in an UHV system on a Cu(100) single crystal, effects due to contamination within the substrate, the chamber, and the gas stream can be minimized.  In addition, the surface normal of the Cu(100) crystal used in this study was oriented within $\pm 0.1 ^\circ$ of the [100] direction to help reduce the influence of step edges on the nucleation of the graphene grains.

\section{Experimental}
\begin{figure}[h]
  \includegraphics[width=6in]{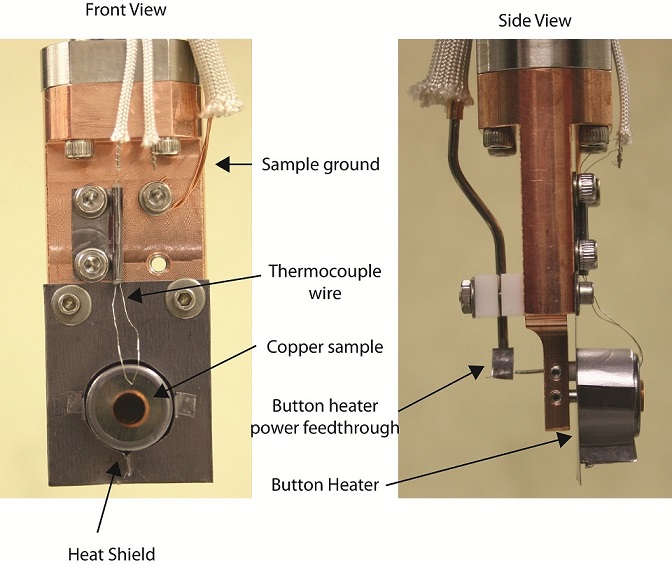}
  \caption{Front view (left) and side view (right) of custom sample holder with Cu(100) crystal mounted on button heater.}
  \label{CustomSampleHolder}
\end{figure}

Graphene growth was carried out on a Cu(100) single-crystal (99.999\% purity) in an UHV chamber with a base pressure of $1 \times 10^{-10}$~Torr.  The Cu(100) crystal was cut in a top-hat design, and the surface was polished to within $\pm 0.1 ^\circ$, which results in an average terrace width of 0.1~$\mu$m.  The substrate was heated with an oxygen series button heater (HeatWave Labs, Inc., P/N 101275-28) that was mounted on a custom-made sample holder, as shown in Figure \ref{CustomSampleHolder}.  The sample holder was attached to a stainless steel dewar that was connected to a differentially pumped rotary motion feedthru that was attached to an \emph{x,y,z} manipulator.  The crystal was held in place with a Ta cap.  A spacer ring made from Mo foil was used to center the crystal in the Ta cap.  Ta foil was mounted on the side and back of the button heater to reduce radiative heat losses.  This allowed annealing of the crystal at temperatures as high as $\sim$1000 $^\circ$C, which is close to the melting point of Cu (1085 $^\circ$C). The front surface of the crystal was left exposed so that the surface could be cleaned by Ar ion sputtering and could be characterized with \emph{in-situ} LEED.  A chromel-alumel thermocouple was spot-welded onto the Mo ring, and was used to calibrate a disappearing-filament pyrometer that was used to measure the temperature of the front surface of the Cu crystal.  LEED was the primary characterization technique used for these experiments and was done \emph{in-situ} with a Princeton Research Instruments (PRI), four-grid, rear-view LEED. SEM was also used to characterize the sample, and those measurements were done \emph{ex-situ} in a LEO 1550 SEM.

The gas doser systems used for the ethylene, oxygen, and argon each consisted of a bakeable variable leak valve, stainless steel tubing, stainless steel regulator, and lecture bottle.  The grade of ethylene was ultra-high purity (99.95\%); the grade of oxygen was Matheson purity (99.997\%); and the grade of the argon was Matheson purity (99.9995\%).   The stainless steel tubing of each doser system was baked into a mini turbo-molecular pump for 24 hours before backfilling with the ethylene, oxygen, or argon.  Ethylene pressures as high as 10 mTorr were used for the growth of the graphene films.  In this pressure range, gas molecules are readily ionized by large electric fields, which will produce a corona discharge.  In addition, ethylene will dissociate when passed across a hot filament.  To prevent inadvertent dissociation of the ethylene, the ion gauge was turned off once the chamber pressure exceeded 10$^{-5}$ Torr.   A UHV capacitive manometer that has a measurement range of $1 \times 10^{-1}$~Torr to $1 \times 10^{-5}$~Torr was used to measure the ethylene pressure during graphene growth.  The capacitive manometer has no high voltages or hot filaments that could result in a corona discharge or the dissociation of the ethylene molecules.  The capacitive manometer is also an absolute pressure gauge, which has the advantage that it does not require calibration for specific gas types.  With this unique UHV-based experimental setup, the growth of graphene by CVD can be performed with precursor pressures as high as 100 mTorr and temperatures up to 1000~$^\circ$C.  Because of the extrene cleanliness of the gas dosing system and the UHV components, the chamber pressure typically returns to the 10$^{-10}$~Torr range within a few minutes of pumping out the precursor gas.  This allows the growth of the graphene films in an environment that minimizes the presence of impurities at the surface of the crystal and within the precursor gas, while allowing the \emph{in-situ} characterization  of the films with LEED.

The Cu(100) single crystal was cleaned by cycles of sputtering with 1~keV Ar ions followed by annealing.  Initial attempts to clean the crystal by sputtering followed by annealing at 650~$^\circ$C resulted in a clean surface after a few sputter-anneal cycles, as determined by LEED patterns with sharp spots, low diffuse background, and the square symmetry expected for the Cu(100) surface.  However, subsequent annealing of the crystal at 900~$^\circ$C resulted in a complex LEED pattern from diffusion of impurities to the surface, presumably S, which is a common impurity in Cu.  In order to sufficiently clean the crystal for high temperature graphene growth, it was necessary to sputter the crystal while annealing at 650~$^\circ$C.  After a few cycles of high temperature sputtering, it was found that annealing in UHV at 900~$^\circ$C resulted in sharp LEED spots with the correct square symmetry associated with the Cu(100) surface. After the bulk impurities were removed from the Cu(100) crystal, a few cycles of room temperature sputtering and annealing were done to ensure a relatively smooth surface for graphene growth since high temperature sputtering can result in surface roughening.  A LEED pattern of the Cu(100) surface after this cleaning procedure is shown in Figure \ref{CleanCu100}a.  A faint ring structure is also observed in the pattern just inside the four primary LEED spots of the Cu(100) substrate.  This is caused by reflection of light from the back of the electron gun and is not related to diffraction from the sample surface. The growth of a chemisorbed oxygen layer on the Cu(100) surface was achieved by dosing molecular oxygen on the surface at a pressure of $1~\times~10^{-6}$~Torr for 5~min (300~L) while maintaining the temperature of the crystal at 300~$^\circ$C.  This resulted in a sharp ($\sqrt{2}\times2\sqrt{2}$)~R45$^\circ$ LEED pattern (Figure \ref{OxygenCu100}a), which corresponds to 0.5 monolayers of chemisorbed oxygen \cite{Wuttig1989103}.  Since the Cu(100) surface has square symmetry and the oxygen reconstruction has rectangular symmetry, two different orientations of the reconstruction are observed in the figure, rotated by 90$^\circ$ with respect to each other.

\section{Results}

The initial attempts to grow graphene on the Cu(100) surface used a technique where the substrate was heated to the growth temperature and the ethylene was then backfilled into the UHV chamber with the gate valves to the ion pump and turbo pump closed.  Growth temperatures of 700~$^\circ$C, 800~$^\circ$C and 900~$^\circ$C were attempted with ethylene pressures ranging from 1 to 10~mTorr. In each attempt, the Cu substrate was held at the growth temperature in ethylene for 10 minutes, then the power to the button heater was ramped down, and the gate valve to the turbo pump was opened. The initial cooling rate was $\sim$70$^\circ$C/min, and the chamber pressure dropped to the low 10$^{-9}$ Torr range within about a minute of opening the turbo pump gate valve.  For each of these growth attempts, no graphene was detected with LEED  after the Cu(100) crystal had cooled to room temperature.  Graphene growth was also attempted with the gate valve to the turbo pump open to produce a gas-flow environment.  This also did not result in graphene growth on the Cu(100).  These results are similar to our previously published attempts to grow graphene on the Cu(111) substrate by heating the crystal and backfilling with ethylene \cite{robinson2012argon}.  The suppression of graphene growth on these surfaces is attributed to the high vapor pressure of Cu at the graphene growth temperature.  For instance, the vapor pressure of the Cu(100) surface is $4 \times 10^{-6}$~Torr at 900~$^\circ$C \cite{vaporpressure}, which results in a sublimation rate of 0.4~ML/s in UHV.

\begin{figure}[h]
  \includegraphics[width=6in]{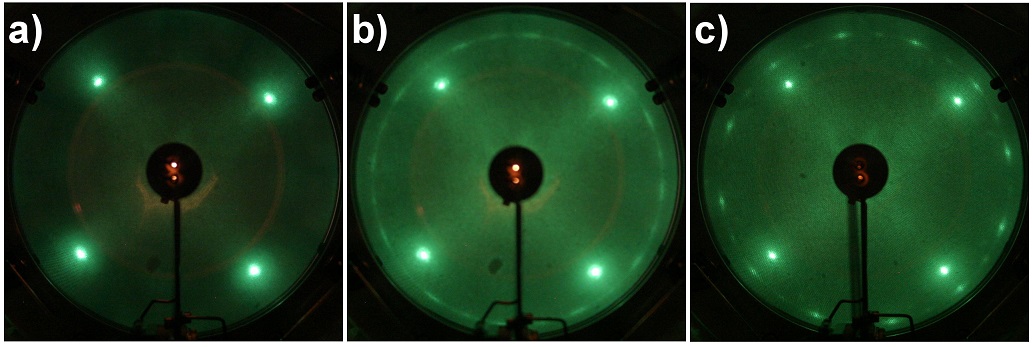}
  \caption{LEED patterns of a) the clean Cu(100) surface, b) the Cu(100) surface following graphene growth in 5~mTorr ethylene for 10~min at 800~$^\circ$C, and c) the Cu(100) surface following graphene growth in 5~mTorr ethylene for 10~min at 900~$^\circ$C.  All patterns taken at 70~eV.}
  \label{CleanCu100}
\end{figure}

A growth technique that involved first backfilling the chamber with ethylene while the crystal is at room temperature, followed by heating the substrate to the growth temperature was then attempted.  For each growth, the crystal was held at the growth temperature for 10~min, after which the gate valve to the turbo pump was opened, and power to the button heater was ramped down.  This technique resulted in graphene formation on the Cu(100) substrate.  A LEED pattern taken after the growth of graphene on the Cu(100) at an ethylene pressure of 5~mTorr and an anneal temperature of 800~$^\circ$C is shown in Figure \ref{CleanCu100}b. The four first-order diffraction spots from the Cu(100) surface can be seen in the figure, as well as a ring just outside the Cu(100) spots with approximately 24 broad regions of increased intensity. This corresponds to graphene domains with four different primary rotational alignments with respect to the Cu(100) substrate lattice. The arcs indicate that the alignment of each graphene domain has considerable rotational disorder.  The same growth conditions were repeated at 900~$^\circ$C and resulted in the corresponding LEED image shown in Figure \ref{CleanCu100}c.  For this growth condition, the graphene has a much higher degree of rotational order, as indicated by the formation of spots instead of broad arcs. The 12 spots correspond to two predominant orientations of the graphene with respect to the Cu(100) substrate.  Each domain corresponds to graphene grains with one of their lattice vectors aligned with one of the Cu(100) surface lattice vectors.  Since the Cu(100) surface has square symmetry but graphene has hexagonal symmetry, this results in a two-domain growth.  In addition, there are faint arcs corresponding to a small amount of graphene that is mis-oriented with respect to the substrate by $\pm$15$^\circ$.

\begin{figure}[h]
  \includegraphics[width=6in]{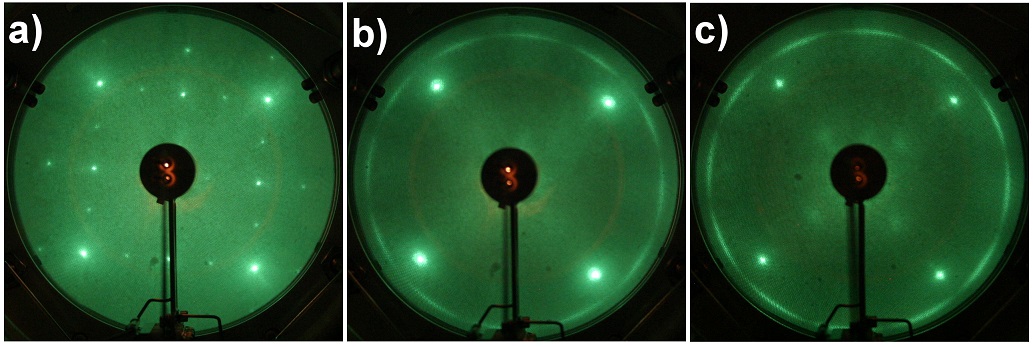}
  \caption{LEED patterns of a) the clean Cu(100) surface dosed with O$_2$ at 300~$^\circ$C and 1~$\times$~10$^{-6}$ Torr for 5~min, b) the oxygen pre-dosed Cu(100) surface following graphene growth in 5~mTorr ethylene for 10 min at 800~$^\circ$C, and c) the oxygen pre-dosed Cu(100) surface following graphene growth in 5~mTorr ethylene for 10~min at 900~$^\circ$C.  All patterns taken at 70~eV.}
  \label{OxygenCu100}
\end{figure}

Once the procedure for growing graphene on the Cu(100) surface was established, the effect that pre-dosing chemisorbed oxygen on the Cu(100) surface has on the growth of graphene was studied. Before each graphene growth was attempted, the Cu(100) surface was cleaned by sputtering and annealing, and a chemisorbed oxygen layer was formed on the Cu(100) surface by dosing oxygen while annealing the crystal at 300~$^\circ$C. The graphene growth procedure used after the chemisorbed oxygen layer was formed was identical to the procedure used for the clean Cu(100) surface.  The LEED patterns for growths done at 800~$^\circ$C and 900~$^\circ$C in 5mTorr of ethylene for 10 min are shown in Figure \ref{OxygenCu100}b and \ref{OxygenCu100}c, respectively. In both cases, a ring structure consisting of 12 broad arcs is observed.  The 12 arcs correspond to graphene with one of its lattice vectors aligning $\pm$15$^\circ$ out of phase with one of the Cu(100) surface lattice vectors.  The primary difference between the LEED patterns for growth at 800~$^\circ$C and 900~$^\circ$C is that the intensity of the arcs after growth at 900~$^\circ$C is stronger than for growth at 800~$^\circ$C.  In addition, the Cu(100) spots are weaker after growth at 900~$^\circ$C, which indicates that the graphene overlayer has a higher coverage at this temperature.

\begin{figure}[h]
  \includegraphics[width=3in]{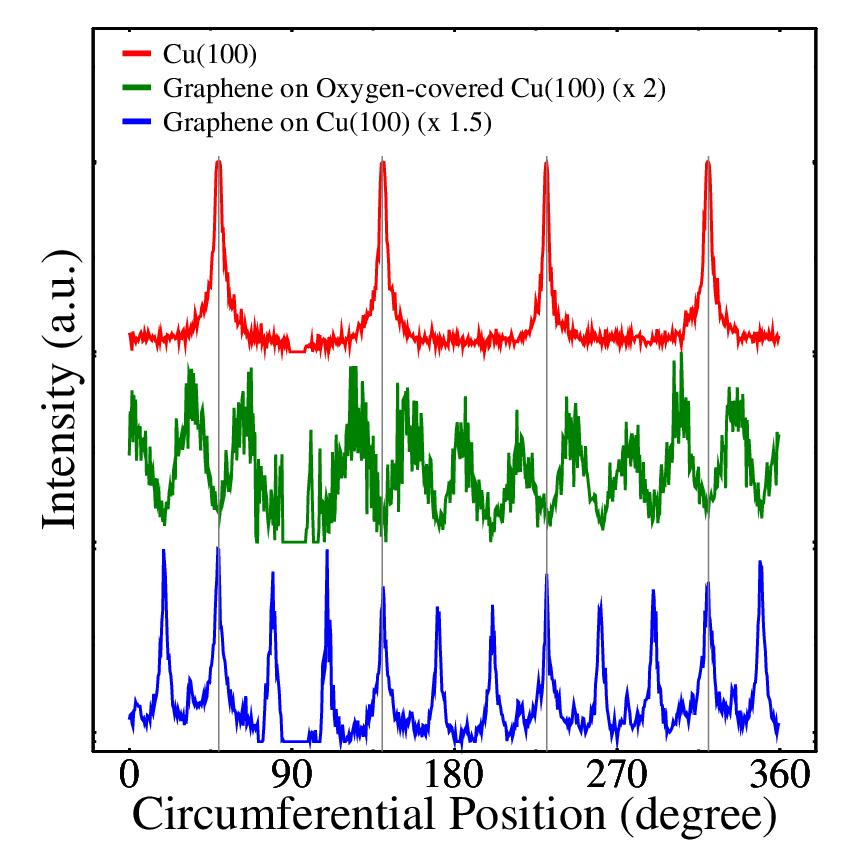}
  \caption{Circumferential intensity profiles for  the clean Cu(100) surface (top), graphene grown on the $\sqrt{2}~\times~2\sqrt{2}$R45$^\circ$ oxygen reconstructed surface (middle), and graphene grown on the clean Cu(100) surface (bottom).  Vertical lines spaced by 90$^\circ$ are drawn to guide the eye.  Intensity profiles offset for clarity.}
  \label{Cu100_bothGraphene}
\end{figure}

For both the graphene grown on the clean Cu(100) surface and the surface with a chemisorbed oxygen layer, circumferential intensity profiles were measured, as seen in Figure~\ref{Cu100_bothGraphene}.  The intensity profile for the clean Cu(100) surface has four sharp peaks, indicating a well ordered surface.  For the lower plot, which is from graphene that was grown on the clean Cu(100) surface, there are 12 primary peaks due to the two different rotational orientations of the 6-fold symmetric hexagonal lattice of the graphene. These peaks are also sharp, indicating a small amount of rotational disorder. In the middle plot, which is from graphene that was grown on the Cu(100) surface that was pre-dosed with chemisorbed oxygen, 12 broad peaks can be seen that are out of phase with the Cu(100) spots.

Since Alstrup \emph{et al}. \cite{Alstrup199295} had observed that pre-dosing oxygen on the Cu(100) surface reduced the activation energy for the dissociation of methane, growths at 700~$^\circ$C for 10 min at an ethylene pressure of 5 mTorr on both the oxygen pre-dosed and clean Cu(100) surfaces were attempted.  Trace amounts of graphene formed on both surfaces, as indicated by the very weak arcs observed in the LEED pattern for growth on the clean Cu(100) surface and a very weak ring structure for the oxygen pre-dosed surface.  Since the LEED intensity associated with the graphene overlayer was approximately the same for both surface preparations, the pre-dosing of oxygen does not seem to significantly lower the activation energy for dissociation of ethylene at the Cu(100) surface.

\begin{figure}[h]
  \includegraphics[width=6in]{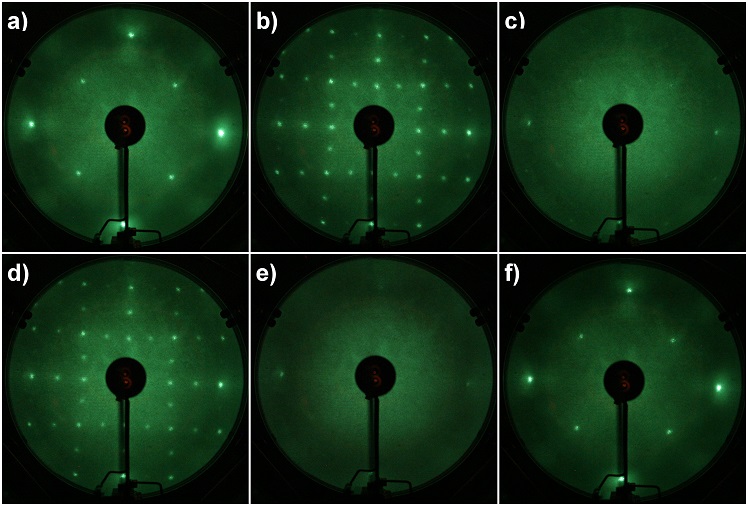}
  \caption{LEED images taken while heating the Cu(100) surface that had been pre-dosed with a chemisorbed oxygen overlayer.  Images were taken for a) the clean Cu(100) surface at RT, b) the surface at RT after dosing with oxygen, c) while annealing at 500~$^\circ$C, d) after cooling back down to 100~$^\circ$C, e) while annealing at 700~$^\circ$C, and f) after cooling back down to RT.  The images were taken with incident electron energy of 112eV.}
  \label{CookAndLook}
\end{figure}

To determine the thermal stability of the chemisorbed oxygen layer, temperature dependent LEED analysis was done on the oxygen pre-dosed Cu(100) surface, as shown in Figure~\ref{CookAndLook}.  A chemisorbed oxygen layer was first adsorbed on the Cu(100) surface, resulting in a sharp ($\sqrt{2}\times2\sqrt{2}$)~R45$^\circ$ LEED pattern (Figure~\ref{CookAndLook}b).  The sample was then heated while simultaneously performing LEED analysis.  At ~$\sim$400~$^\circ$C, the LEED pattern was observed to convert from a ($\sqrt{2}\times2\sqrt{2}$)~R45$^\circ$ pattern to a ($\sqrt{2}\times\sqrt{2}$)~R45$^\circ$ pattern.  At 500~$^\circ$C, the LEED spots corresponding to the oxygen overlayer had almost completely disappeared and the Cu(100) substrate spots were very weak (Figure~\ref{CookAndLook}c).  However, upon cooling the sample back down to 100~$^\circ$C, the sharp ($\sqrt{2}\times2\sqrt{2}$)~R45$^\circ$ LEED pattern returned (Figure~\ref{CookAndLook}d).   This indicates that the chemisorbed oxygen overlayer undergoes a surface melting transition at $\sim$500~$^\circ$C and then reorders in the ($\sqrt{2}\times2\sqrt{2}$)~R45$^\circ$ overlayer structure upon cooling.  The heating sequence was then repeated, but to a temperature of 600~$^\circ$C.  At 600~$^\circ$C, the LEED spots corresponding to the oxygen overlayer had completely disappeared and the spots corresponding to the Cu(100) substrate were very weak.  Upon cooling back down to 100~$^\circ$C, the sharp ($\sqrt{2}\times2\sqrt{2}$)~R45$^\circ$ LEED pattern returned, but with weaker intensity than was observed after the first anneal at 500~$^\circ$C.  This indicates that there was a loss of some of the oxygen atoms from the surface.  The heating sequence was then repeated again, but to a temperature of 700~$^\circ$C.  At this temperature, the LEED spots corresponding to the oxygen overlayer and the first order Cu(100) spots had completely disappeared (Figure~\ref{CookAndLook}e).  Upon cooling back down to RT, the LEED pattern of the clean Cu(100) surface was observed (Figure~\ref{CookAndLook}f).  Therefore, annealing at 700~$^\circ$C was sufficient to remove the oxygen from the surface region of the Cu(100) crystal.

\begin{figure}[h]
  \includegraphics[width=3in]{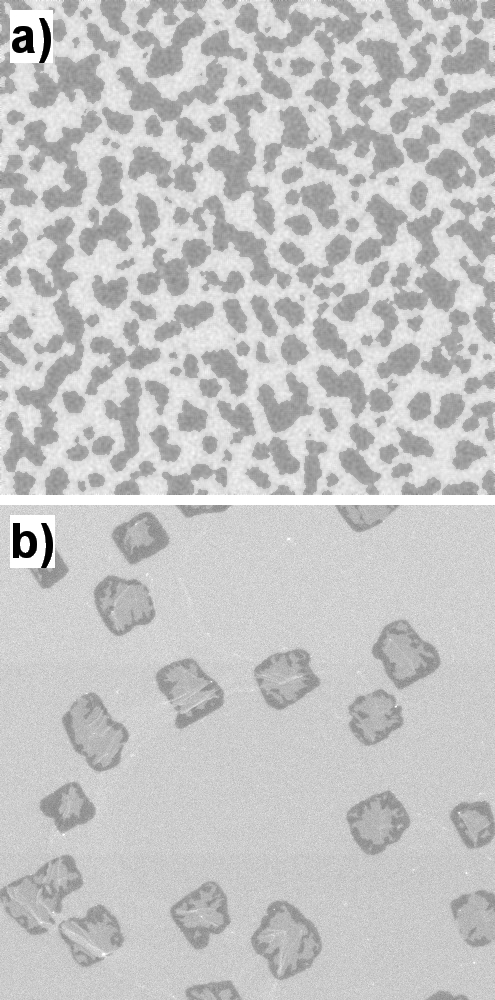}
  \caption{SEM image of graphene grown on a the clean Cu(100) surface (5~$\mu$m~$\times$~5~$\mu$m) and b) the oxygen pre-dosed Cu(100) surface (12~$\mu$m~$\times$~12~$\mu$m). Images filtered with ImageJ.}
  \label{SEM}
\end{figure}

To determine the growth morphology of the graphene on both the clean and oxygen pre-dosed Cu(100) surfaces, \emph{ex-situ} SEM analysis was performed.  For each surface, the graphene overlayer was grown by backfilling the UHV chamber with 5~mTorr of ethylene, followed by ramping the temperature to  900~$^\circ$C and annealing for 5~min.  After venting, the crystal was transferred to the SEM so that the surface coverage and nucleation rate could be determined. The LEED images that were taken before venting the chamber indicate that a well-ordered, two-domain, epitaxial graphene overlayer had formed for growth on the clean Cu(100) surface and that a two-domain overlayer with considerable rotational disorder had formed for growth on the oxygen pre-dosed surface.  SEM images of the graphene overlayers are shown in Figure \ref{SEM}. For the graphene grown on the clean Cu(100) surface, the graphene regions appear as dark patches (Figure \ref{SEM}a).  For the graphene grown on the oxygen pre-dosed surface, the graphene regions have a dark patch at the outer edge of the graphene island (Figure \ref{SEM}b).  The dark constrast is attributed to oxidation of the Cu atoms below the graphene.   The lateral size of the graphene islands grown on the clean surface is $\sim$0.1~$\mu$m, whereas the lateral size of the graphene islands grown on the oxygen pre-dosed surface is $\sim$1.5~$\mu$.  Since the islands on the oxygen pre-dosed surface are considerably larger than the islands grown on the clean surface, the Cu oxidation is only present under the perimeter of the islands grown on the oxygen pre-dosed surface.  For both substrate preparations, graphene islands formed randomly on the substrate surface.  For growth on the clean Cu(100) surface, the islands were irregularly shaped.  The graphene coverage is estimated to be 45\% and the nucleation rate to be $\sim$3~islands per $\mu$m$^2$.  The exact nucleation density is difficult to estimate due to the coalescence of some of the islands.  For growth on the oxygen pre-dosed surface, the islands are somewhat square in shape.  The graphene coverage is 20\%, and the nucleation rate is $\sim$0.1~islands per $\mu$m$^2$.  Therefore, the much larger size of the graphene islands on the oxygen pre-dosed surface is a result of the much lower nucleation rate.  The lower absolute coverage probably results from the reaction of some of the surface oxygen with carbon atoms to form volatile CO and CO$_2$.

\section{Discussion}
The observation that graphene would not grow when exposing the Cu crystal to ethylene after heating it to the growth temperature but could be grown when the substrate temperature was ramped from RT to the growth temperature in an ethylene atmosphere can be explained by the temperature dependence of the ethylene dissociation process.  If the dissociation of the ethylene molecules begins in the temperature range where the the vapor pressure of the Cu is relatively low, a non-graphitic carbon layer will form on the surface before Cu sublimation becomes appreciable.  Once the temperature of the substrate reaches a point where the Cu vapor pressure is high, the carbon on the surface will act as a diffusion barrier for the subliming Cu atoms.  Interestingly, we reported in a previous publication that when graphene growth was attempted on the Cu(111) surface by heating the crystal from RT to the growth temperature in ethylene, only trace amounts of graphene could be formed \cite{robinson2012argon}.  This indicates that the catalytic activity of the Cu(100) surface is higher than the Cu(111) surface for the decomposition of ethylene.  Because the atoms at the (100) surface of FCC metals have a lower coordination than the atoms at the close-packed (111) surface, the (100) surface is generally more reactive than the (111) surface for most metals.

\begin{figure}[h]
    \includegraphics[width=5in]{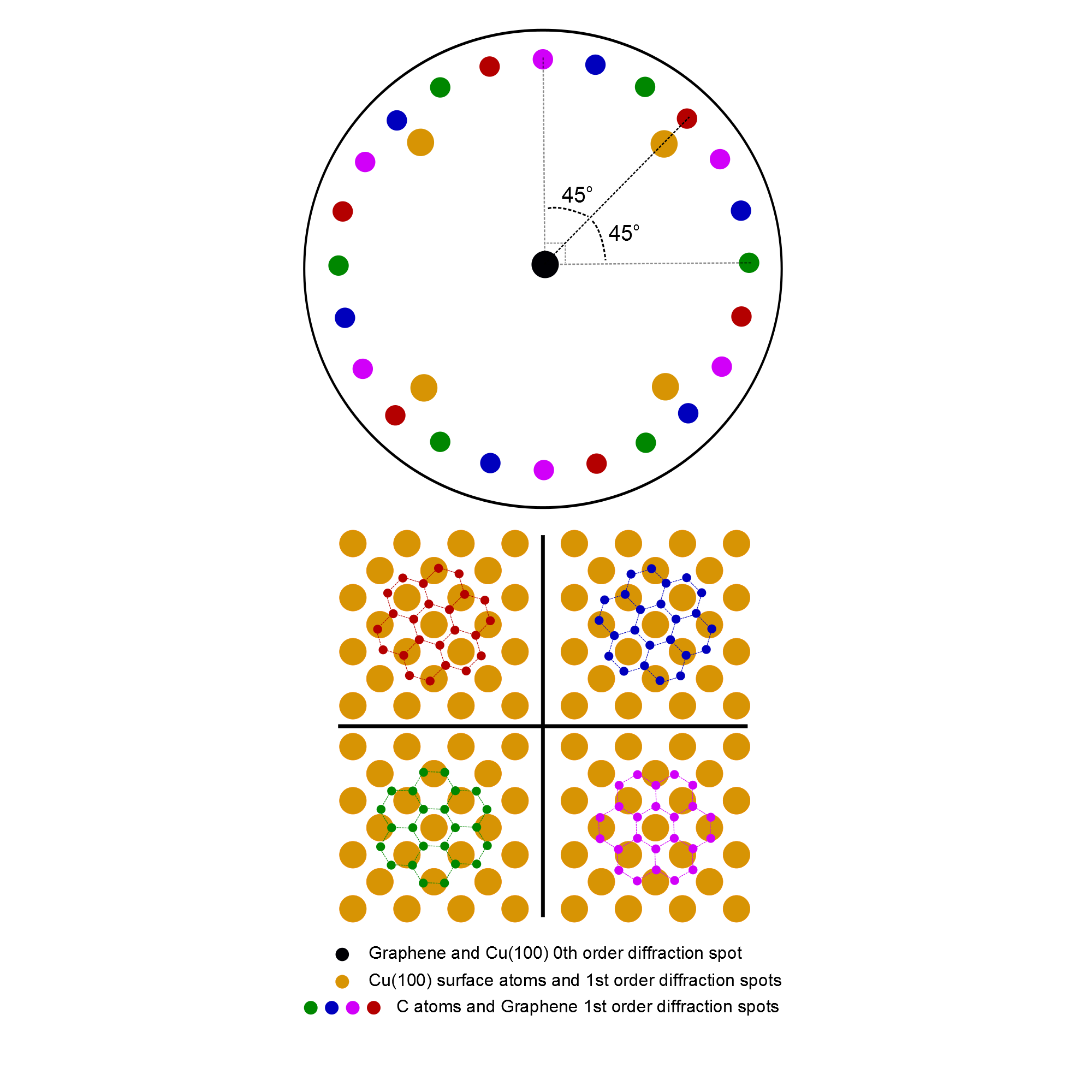}
    \caption{Simulated LEED image for graphene grains with four different rotational orientations with respect to the Cu(100) surface at an electron energy of 70~eV.  Nucleation of a graphene island with every third C-C bond in the direction of the Cu-Cu nearest-neighbor direction results in a diffraction pattern with two graphene spots rotationally aligned with two of the Cu(100) spots.  Nucleation of a graphene grain with every third C-C bond in the direction of the Cu-Cu next-nearest-neighbor direction results in a diffraction pattern with two graphene spots rotated 45$^\circ$ with respect to two of the Cu(100) spots.}
    \label{model}
\end{figure}

For graphene growth on the clean Cu(100) substrate, four rotational orientations of the graphene grains with respect to the substrate are observed for growth at both 700~$^\circ$C and 800~$^\circ$C.  A schematic of the four different rotational orientations of graphene and the corresponding LEED pattern are shown in Figure \ref{model}.  As the growth temperature is increased to 900~$^\circ$C, two domains are observed: each with one of the reciprocal lattice vectors of the hexagonal lattice of the graphene aligned with one of the reciprocal lattice vectors of the square lattice of the Cu(100) surface.  The observation of four domains at lower temperatures indicates that there are two nucleation orientations that are almost equivalent in energy.  As the temperature is increased, the thermal energy will eventually become sufficient for all of the carbon atoms to access the lowest energy orientation (\emph{i.e.}, the orientation where one of the graphene lattice vectors is aligned with one of the surface lattice vectors of the Cu(100)).  Two rotational orientations of the graphene grains are also observed for growth on the oxygen pre-dosed Cu(100) surface.  However, each orientation has one of its reciprocal lattice vectors weakly aligned $\pm$45$^\circ$ with respect to one of the Cu(100) reciprocal lattice vectors.  Because of the six-fold symmetry of the hexagonal lattice, this results in the observed $\pm$15$^\circ$ rotation graphene diffraction arcs with respect to the Cu(100) diffraction spots (Figure \ref{OxygenCu100}c).

The change in preferred alignment and the increase in rotational disorder of the graphene grains for growth on the oxygen predosed Cu(100) surface indicates that during the initial nucleation of the graphene, oxygen is suppressing the formation of grains that have one of their lattice vectors aligned with one of the lattice vectors of the Cu(100) surface.  In other words, the presence of oxygen at the surface is most likely blocking the nucleation of graphene grains on the Cu(100) terrace sites.  Since the temperature dependent LEED analysis provides evidence that most of the oxygen has left the surface region of the Cu(100) substrate by $\sim$700~$^\circ$C, this would imply that the initial nucleation of graphene is occurring below 700~$^\circ$C.  This is further corroborated by the observation that trace amounts of graphene formed on both the clean and oxygen pre-dosed surfaces when heated to 700~$^\circ$C in ethylene.  For graphene growth on the clean Cu(100) surface, an increase in graphene grains aligned with the Cu(100) substrate was observed as the growth temperature was raised from 800~$^\circ$C to 900~$^\circ$C.  This was not observed for growth on the oxygen pre-dosed surface.  This is probably because the size of the graphene islands for the oxygen pre-dosed surface are about an order of magnitude larger than for the clean Cu(100) surface.  Therefore, the islands may be too large to rotate into one of the two equivalent low energy orientations at elevated temperatures.  It may also be that the graphene islands that are nucleating on the oxygen pre-dosed surface are composed of grains with multiple orientations.  Therefore, these graphene islands will not have a preferred orientation with respect to the Cu(100) substrate.  Since the electron gun of our conventional LEED has a diameter of $\sim$1~mm, it is probing the crystallographic orientation of several thousand graphene islands simultaneously.  To probe the crystallographic orientation of each individual graphene island, a $\mu$-LEED analysis would need to be done using LEEM.

The role that oxygen plays in graphene growth is important because most CVD reactors operate in either the low pressure or atmospheric pressure regime where some oxygen contamination will be present.  Most groups use a flow of hydrogen in their reactor to reduce the copper oxide that is in the bulk or forms at the surface of the substrate.  The reduction process will convert the copper oxide into metallic Cu and water vapor. However, as shown above, even sub-monolayer oxygen contamination on the surface of the Cu(100) substrate can drastically affect the subsequent graphene growth. For instance, our results show an increase in rotational disorder, a change in the preferred orientation of the graphene grains, and a reduction in the nucleation rate of the graphene islands for growth on a substrate with a half monolayer of oxygen atoms chemisorbed to the surface.  These results could help explain why different groups have measured large differences in the transport properties for graphene films grown on Cu foil substrates even though the reported growth conditions are similar.

Since a minimum of two rotational orientations are expected for graphene growth on the Cu(100) surface, grain boundaries will always be present once the graphene islands that have nucleated on the surface to coalesce into a single monolayer film.  For perfect two-domain epitaxial growth on the Cu(100) surface, the grains are rotated 90$^\circ$ with respect to each other.  Yazyev and Louie have predicted that the reflection probability of carriers in graphene depends strongly on the misorientation angle between adjacent grains \cite{5388152720101001}.  For a 90$^\circ$ mis-orientation, a transport gap of 1.1 eV is predicted.  Therefore, this particular orientation of adjacent grains is expected to result in a high probability for reflection of charge carriers at the grain boundary.  Since the graphene grains grown on a Cu(100) surface that has been pre-dosed with chemisorbed oxygen have a much larger distribution of misalignment angles, a wide range of scattering probabilities at grain boundaries will result.  Therefore, weak rotational alignment of the graphene grains with respect to the substrate may produce better transport properties than a well-aligned two-domain epitaxial overlayer.  The actual transport properties measured for a particular graphene film will depend on both the mis-orientation angle(s) of the grains and the average size of the graphene grains.

\section{Conclusions}
Our results show that a well-ordered two-domain epitaxial graphene film can be grown on the clean Cu(100) surface at 900~$^\circ$C.  On the other hand, the presence of a chemisorbed oxygen layer on the Cu(100) surface before graphene growth results in graphene overlayers with considerable rotational disorder.  The difference in growth morphology of the graphene films grown on the clean and oxygen pre-dosed Cu(100) surfaces is attributed to the interaction of oxygen with the ethylene molecules during the initial nucleation of graphene islands.  Because Cu foils are one of the most common substrates for large area graphene growth and they typically recrystallize with a (100) texture during graphene growth, our results for the growth of graphene on both the clean and oxygen pre-dosed Cu(100) surfaces could be used to develop techniques for growing graphene films with better transport properties.

\acknowledgement
This research project was supported by the National Science Foundation (DMR-1006411).  P. T. would like to thank NSF for her financial support, and T. R. M. would like to thank SEMATECH for his financial support. Z. R. R. would like to thank the American Society for Engineering Education for his post-doctoral support.

\bibliographystyle{achemso}

\bibliography{bibliography}

\pagebreak

\pagebreak

\pagebreak

\pagebreak

\pagebreak

\pagebreak

\pagebreak

\end{document}